# MESOSCOPIC COLLECTIVE ACTIVITY IN EXCITATORY NEURAL FIELDS: CROSS-FREQUENCY COUPLING

Abstract. In the brain, cross-frequency coupling has been hypothesized to result from the activity of specialized microcircuits. For example, theta-gamma coupling is assumed to be generated by specialized cell pairs (PING and ING mechanisms), or special cells (e.g., fast bursting neurons). However, this implies that the generating mechanisms is uniquely specific to the brain. In fact, cross-scale coupling is a phenomenon encountered in the physics of all large, multi-scale systems: phase and amplitude correlations between components of different scales arise as a result of nonlinear interaction. Because the brain is a multi-scale system too, a similar mechanism *must* be active in the brain. Here, we represent brain activity as a superposition of nonlinearly interacting patterns of spatio-temporal activity (collective activity), supported by populations of neurons. Cross-frequency coupling is a direct consequence of the nonlinear interactions, and does not require specialized cells or cell pairs. It is therefore universal, and must be active in neural fields of any composition. To emphasize this, we demonstrate the phenomenon in excitatory fields. While there is no doubt that specialized cells play a role in theta-gamma coupling, our results suggest that the coupling mechanism is at the same time simpler and richer: simpler because it involves the universal principle of nonlinearity; richer, because nonlinearity of collective activity is likely modulated by specialized-cell populations in ways to be yet understood.

Short running title: Collective Neural Activity: Cross-frequency Coupling

Authors: Y. Qin[1] and A. Sheremet[1,2]

Affiliations: 1. Engineering School of Sustainable Infrastructure and Environment, University of Florida, Gainesville, FL. 32611.
2. McKnight Brain Institute, Department of Neuroscience, University of Florida, Gainesville, FL. 32610.

Correspondence: Alex Sheremet, email: alex.sheremet@essie.ufl.edu

Competing Interests: The authors declare that they have no competing interests.

Data Availability: No laboratory data was used in the research presented in this manuscript.

Funding information: This work was supported by the McKnight Brain Research Foundation, National Institute on Aging, grant number AG055544, and National Institute of Mental Health, grant number MH109548.

Key words: Neural field equations · Multi-scale modeling · Population dynamics · Stability and bifurcation · Cross-frequency Coupling · Energy cascade

MSC codes: 35Q92 · 92B20 · 92C20

## Contents

## 1. Introduction

Cross-frequency coupling may be defined as a statistical correlation between characteristic parameters of two LFP frequencies. The most studied example is probably the coupling of the theta and gamma rhythms [e.g., Soltesz and Deschênes, 1993], but the family of cross-frequency coupling phenomena could be extended to include also of non-periodic and transient oscillations, such as sharp-waves/ripples, and slow-oscillations/spindles. Theta-gamma coupling was identified in rat hippocampus [e.g., Soltesz and Deschênes, 1993, Bragin et al., 1995, Rosenblum et al., 2000, Colgin et al., 2009, Lisman, 2005, Belluscio et al., 2012, Colgin, 2014, Pernía-Andrade and Jonas, 2014], in the visual cortex of primates [Whittingstall and Logothetis, 2009, Mazzoni et al., 2011], and in humans [Canolty et al., 2006]. Fundamental contributions to understanding the sharp-waves/ripple complex include Buzsáki et al. [1992], Skaggs et al. [2007], Bragin et al. [1999], Buzsaki [2006], Clemens et al. [2011], Buzsáki and da Silva [2012]; the coupling of low-oscillations and spindles was investigated by Marshall et al. [2006], Rasch and Born [2013], Staresina et al. [2015], Ladenbauer et al. [2017]. The published literature on the topic is vast, and it is hard to do justice to all relevant contributions in the space of this introduction. Jensen and Colgin [2007] hypothesize that "given that this coupling reflects a specific interplay between *large ensembles of neurons*[1], (cross-frequency coupling) it is likely to have profound implications for neuronal processing", a suggestion supported by further research, which indicates it may play a role in the formation of short-term memories [Colgin et al., 2009, Lisman and Jensen, 2013, Pernía-Andrade and Jonas, 2014, Lega et al., 2014, Bergmann and Born, 2018].

Theta-gamma coupling is often understood as a consequence of mechanisms for gamma generation. In networks with inhibition, interneurons are assumed to play an essential role in gamma generation. The PING mechanism [Bartos et al., 2007, Tiesinga and Sejnowski, 2009, Buzsáki and Wang, 2012] pairs excitatory and inhibitory cells: the reactivation delay of excitatory cells after inhibition by interneurons then determines the oscillation period of the pair and also introduces a phase lag between excitatory and inhibitory activity [Buzsáki and Wang, 2012]. The ING mechanism for inhibitory networks is based on the balance between excitatory drive and inhibitory post-synaptic potentials [Whittington et al., 1995, Buzsáki and Wang, 2012]. In the absence of inhibition, gamma may be due to special fast-bursting neurons [Calvin and Sypert, 1976, Gray and McCormick, 1996], or to gap junctions in the axons of pyramidal cells [Traub et al., 1999a,b, 2000]. In general, the coupling phenomenon occurs at cell-scale, using either pairs of specialized cells, or specific cell features, and becomes detectable at mesoscale through synchronization [e.g., Kuramoto, 1975, Strogatz, 2000].

All above mechanisms identify *cross-frequency coupling as phenomenon strictly specific to brain activity*. In fact, cross-frequency coupling is ubiquitous in multi-scale physical systems that comprise a large number of nonlinearly interacting elements, such as water waves, hydrodynamic vortices, plasma physics, nonlinear optics, Bose-Einstein condensation, coagulation-fragmentation processes, and others [e.g., Zakharov et al., 1992b, Nazarenko, 2011, L'vov, 1998]. In these systems, the coupling mechanism is nonlinearity: interacting components exchange energy and create cross-scale phase and amplitude coupling. A demonstration of nonlinear generation of cross-scale[2] coupling is given in Sheremet et al. [2019], equations 17, based on the canonical three-wave equation [Craick, 1985, Rabinovich and Trubetskov, 1989]. It seems then reasonable to ask if this type of cross-frequency coupling is also active in the brain. By analogy with the physical systems mentioned above, the coupling mechanism would be provided by nonlinear interactions of spatio-temporal activity patterns formed by populations of neurons, or collective activity (see companion paper, Qin and Sheremet, 2022).

[1] Our emphasis.

Here, we demonstrate the role of nonlinearity in coupling collective activity across scales. Because the essential ingredient of this mechanism is nonlinear interactions, specialized cells are not needed to create coupling. This raises an intriguing question: *can nonlinearity of collective-activity give rise to gamma-like oscillations without the help of inhibitory neurons or rhythmic bursting cells?* To answer this question we use the model for mesoscopic activity developed in companion paper, Qin and Sheremet [2022], which is specifically focused on excitatory neural fields.

In LFP recordings, theta-gamma coupling has been quantified using all conceivable correlation measures: phase-phase, phase-frequency, phase-amplitude, and amplitude-amplitude [e.g., Jensen and Colgin, 2007]; frequency-frequency [e.g., Witte et al., 2008]; and amplitude-frequency [e.g., Chehelcheraghi et al., 2017]. While these methods have the advantage of simplicity, they also usually carry strong restrictive assumption: the target components that are supposed to be coupled are defined a priori, and are assumed monochromatic, characterized by a single frequency and phase. In reality, the rhythms occupy frequency bands of significant width (25% to 0.5% of the center frequency, e.g., Sheremet et al., 2019), and coupling occurs at various degrees across the entire band. To eliminate these restrictions, we quantified coupling here using bispectral analysis, which is free of a priori assumptions (it provides a map of coupling intensity across all frequencies) and does not require elaborate data preparation that might destroy or distort information (e.g., filtering, whitening, and others – see the in depth discussion in Kovach et al., 2018).

The interpretation of the bispectrum is straightforward: in the absence of phase correlations the bispectrum is statistically zero; hence, statistically significant non-zero bispectral peaks indicate cross-frequency coupling (see the appendix; also Sheremet et al. 2016, 2020). Figures 1-2, [e.g., Sheremet et al., 2016, 2019, 2020] show examples of bispectra of rat hippocampal LFP corresponding to low and high rat speed, respectively. At low rat speed (figure 1), bispectra are statistically zero, with the exception of two nonzero regions that correspond to weak coupling between theta and its second harmonic and theta and gamma. At high rat speed (figure 2)one related to the coupling between theta and its harmonics (16 Hz to 60 Hz; e.g., Sheremet et al., 2016); the other to the coupling between theta and a wide gamma band of frequencies above theta harmonics (between 60 and 120 Hz).

Section 2 discusses the mechanism of the cross-scale coupling on a generic model. The basic elements of the neural field model derived by Qin and Sheremet [2022] are briefly reviewed in subsection 3.1. The rest of section 3 investigate cross-frequency coupling in the neural field model, and discusses various possible dynamical regimes. We summarize and discuss the results in section 4. The appendix provides an informal short introduction to bispectral analysis.

[2]Because nonlinearity couples all characteristic parameters describing the interacting elements, phases as well as amplitudes, “cross-scale” coupling is a slightly more general term than “cross-frequency” and will be preferred below.

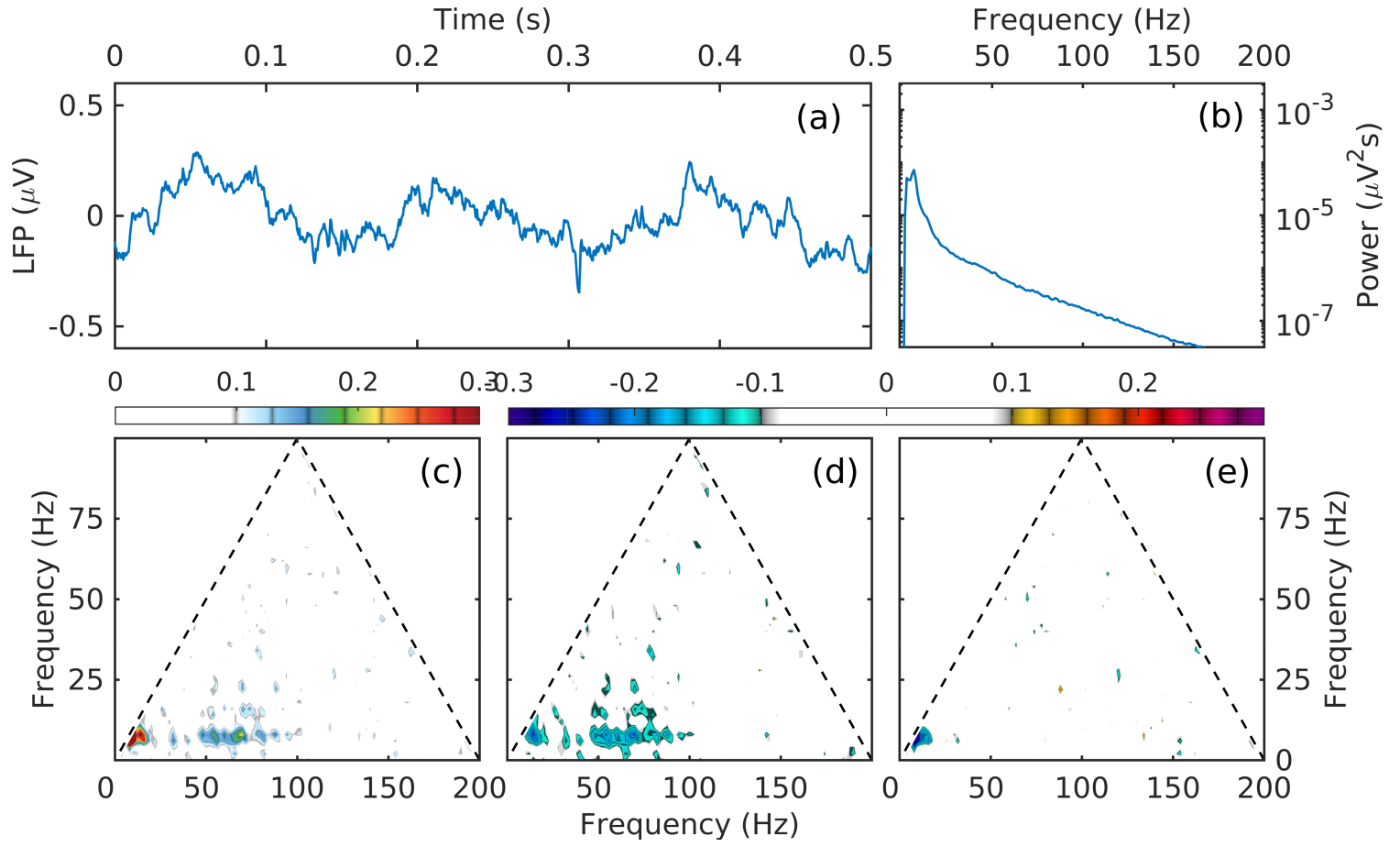

FIGURE 1. Spectrum and bispectrum elements of LFP in the rat hippocampal lacunosum moleculare at low speed (<10 cm/s; see Sheremet et al., 2020). a) A half-second sample of an LFP trace; (b) spectral density of LFP power; (c) modulus of bispectrum, an overall measure of cross-frequency phase coupling; d) real part of the bispectrum, related to the skewness of the LFP trace; and e) imaginary part of the bispectrum, related to the asymmetry of the LFP trace.

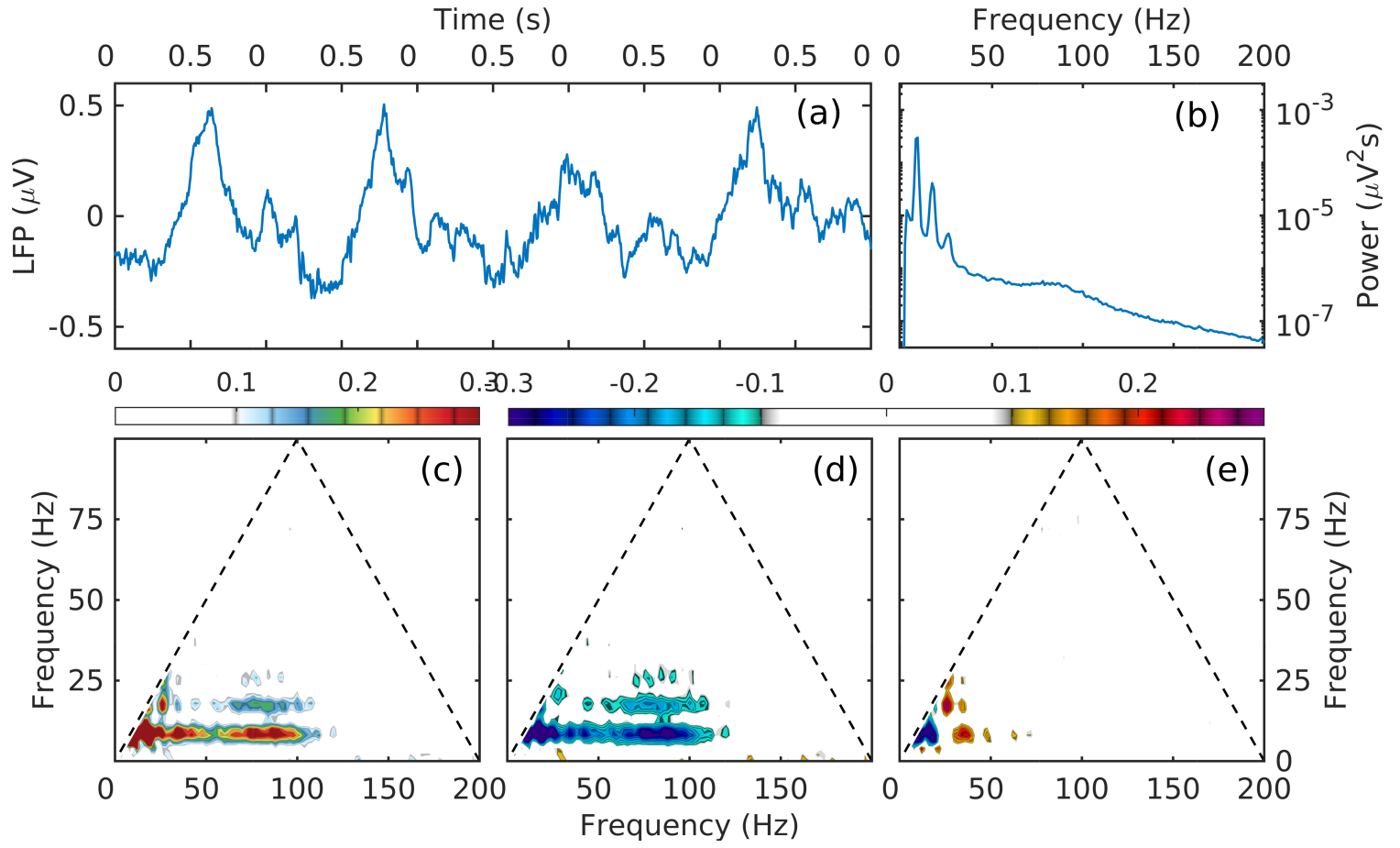

FIGURE 2. Spectrum and bispectrum elements of LFP in the rat hippocampal lacunosum moleculare at high speed (>30 cm/s; same as figure 1)

## 2. QUASISTATIC EVOLUTION IN A GENERIC MODEL

Reduced to its abstract characteristics, the theta-gamma coupling could be described as the interaction of two collective activity oscillations that have widely separated temporal scales (or frequencies), and widely different power, with the larger scale (low frequency) oscillation several orders of magnitude larger than the small-scale oscillation. For example, in rats, the center frequencies for theta and gamma are approximately one order of magnitude apart, 8 Hz and 100 Hz, respectively [Whittington et al., 1995, Tiesinga and Sejnowski, 2009, Buzsáki and Wang, 2012]); in lacunosum moleculare recordings theta power is two orders of magnitude larger than gamma power [e.g., Sheremet et al., 2019]. Below, we will use frequency to quantify scale (e.g., low frequency means large scale). For brevity, we refer to the larger scale component as "theta" and the smaller scale one as "gamma". The reader may think of theta as a global scale oscillation, while gamma is mesoscopic (Muller et al., 2018; see also the introductory discussion in Qin and Sheremet, 2022).

To investigate their coupling, we use the model developed by Qin and Sheremet, 2022. The interaction between two components of well separated (low and high) frequencies is a typical, well studied scenario nonlinear systems (e.g., Yajima and Oikawa, 1976, Benney, 1977, Djordjevic and Redekopp, 1977, Newell, 1978, and many others, see also Caso-Huerta et al., 2021). While such systems are in general integrable, and may be solved using the inverse scattering transform, a full analytical treatment is beyond the scope of this study. However, the power mismatch allows to simplify the equations and gain some qualitative insight into theta-gamma coupled dynamics. Because theta is much more powerful than gamma, the effect of gamma on theta is negligible, while gamma is entirely driven by theta.

To demonstrate the universality of the coupling effect of nonlinear interactions, we first illustrate the simplification on a formal, generic model model.

2.1. **Generic model.** Assume that the governing equation has the general form

$$\frac{dp}{dt} + \mathcal{D}(Q)p = \mathcal{N}\left(p^2\right) \tag{1}$$

where $p$ is a vector of the state variables of the physical system, $\mathcal{D}$ is a linear operator depending on the forcing $Q$, and $\mathcal{N}$ is a nonlinear function of powers of $p$, of which we retain here only the quadratic term (leading order, if nonlinearity is weak).

Following standard analysis of system equilibria and their stability, let $p_0$ be an equilibrium state under the constant forcing $Q_0$, defined by $\frac{d}{dt}p_0 = 0$. For small perturbations of the equilibrium state, expanding the state variable as $p = p_0 + \delta p_1$, where $\delta \ll 1$ is a small parameter, substituting in to equation (1) and separating powers of $\delta$ yields

$$\left(\frac{d}{dt} + \mathcal{D}(Q_0)\right)p_0 = \mathcal{N}\left(p_0^2\right), \tag{2}$$

$$\left(\frac{d}{dt} + \mathcal{D}(Q_0)\right)p_1 = \mathcal{N}(p_0 p_1) + O\left(\delta^2\right). \tag{3}$$

Setting $\frac{d}{dt}p_0 = 0$ reduces equation (2) to an algebraic equation whose solutions are the equilibrium states. In equation (3), the right-hand side is the same as in equation for $p_0$; the nonlinear term, however, is now linearized, and acts as a forcing induced by $p_0$. Noting that the term $\mathcal{N}(p_0 p_1)$ is in fact linear in $p_1$., equation (3) may be written as

$$\left(\frac{d}{dt} + \mathcal{L}(Q_0, p_0)\right)p_1 = 0, \text{ where } \mathcal{L}(Q_0, p_0) = \mathcal{D}(Q_0) - \frac{\partial \mathcal{N}}{\partial p_1}(p_0). \tag{4}$$

The solution of equation (4) in matrix exponential form [Bellman, 1970] is

$$p_1(t) = p_1(0) + e^{\int_0^t \mathcal{L}(Q_0, p_0)ds} A = p(0) + \sum_n A_n \xi_n e^{\sigma_n t} \tag{5}$$

where $A = (A_1, A_2, \cdots)$ is a vector of arbitrary amplitudes, and $\sigma_n$ and $\xi_n$ are eigenvalues and eigenvectors of matrix $\frac{d\mathcal{D}}{dp}$, where $n = 1, 2, \cdots$ counts the eigenvalues. Imaginary eigenvalues correspond to oscillations, and real ones represent exponential decay or growth. The equilibrium is neutrally stable in the subspace spanned by $\xi_n$ if $\Re\{\sigma_n\} = 0$, stable if $\Re\{\sigma_n\} < 0$, and unstable if $\Re\{\sigma_n\} > 0$. If $\Im\{\sigma_n\} \neq 0$, it represents the frequency of oscillation.

2.2. **Quasistatic evolution.** Assume now that the external forcing varies on a characteristic time scale $O\left(\delta\,|\sigma|^{-1}\right)$, much larger than the characteristic time scale $|\sigma|^{-1}$ of the perturbation $p_1$. Introducing a "slow" time $T = \delta t$, and a "fast" time $t$, formally independent, we set $Q = Q(T)$ and expand, as before,

$$p(t, T) = p_0(T) + \delta p_1(t, T). \tag{6}$$

Substituting the expansion (6) into equation (1) obtains

$$\delta \frac{dp_0}{dT} + \delta \frac{dp_1}{dt} + \mathcal{D}(Q)p_0 + \delta \mathcal{D}(Q)p_1 = \mathcal{N}\left(p_0^2\right) + \delta \mathcal{N}(p_0 p_1) + O\left(\delta^2\right). \tag{7}$$

The equations corresponding to the first two orders of magnitude in $\delta$ are similar to equations (2-3), but with an additional slow time dependence

$$\mathcal{D}\left(Q(T)\right) p_0 - \mathcal{N}\left(p_0^2\right) = 0, \tag{8}$$

$$\left[\frac{d}{dt} + \mathcal{D}\left(Q(T)\right) - \frac{\partial \mathcal{N}}{\partial p_1}(p_0)\right] p_1 = -\frac{dp_0}{dt}, \tag{9}$$

where we removed the distinction between slow and fast times. If the matrices $\mathcal{L}\left(p_0(t); Q(t)\right)$ commute for $t_1 \neq t_2$, the solution of equation (9) in the matrix-exponential form is

$$p_1(t) = e^{\int_0^t \mathcal{L}(s)ds} p_1(0) - e^{\int_0^t \mathcal{L}(s)ds} \int_0^t e^{-\int_0^s \mathcal{L}(u)du} \frac{dp_0}{d\tau} ds, \tag{10}$$

were $\mathcal{L}(t) = \mathcal{L}\left(p_0(t); Q(t)\right)$. Equation (8) describes process $p_0$ as a sequence in time of equilibrium states. The approach used here to linearize the governing equations (11) is therefore a quasistatic evolution approximation. Equation (9) represents process $p_1$ as fast transient process through which the system adjusts to the next equilibrium state.

Equations (8-9) represent a forcing "cascade" from external forcing to the large scale, and finally to the small scale component. Equation (8) includes only the forcing $Q$ and the the large scale component $p_0$: theta arises as a direct response to large-scale external forcing $Q = Q(T)$. In contrast, equation (9) describes the evolution of the smaller scale component $p_1$ (gamma) as a result of the joint forcing by external forcing $Q$, and theta, which appears through the nonlinear term $\mathcal{N}(p_0 p_1)$.

The nature of $p_1$, whether oscillatory or decay, is determined characteristics of the equilibrium states visited during the evolution (term $\mathcal{D}$), modulated by nonlinear forcing. Mesoscale processes intensify as system energy (defined by $p_0$) grows, and at high energy nonlinear forcing dominates. In this representation, mesoscale processes are fast but small eigen-oscillations generated by large-scale forcing, itself a direct response of the system to large scale external input. However, the quasistatic evolution is valid only if it does not break the ordering postulated by equation (6). In other words, it is valid for stable or near neutrally stable equilibrium states, say $\Re\{\sigma_n\} < |\Im\{\sigma_n\}|$, but fails if the evolution runs through strongly unstable equilibrium states ($\Re\{\sigma_n\} > 0$ for some $n$), because the solution (10) grows exponentially in time, breaking the expansion (6) over the fast time scale. Therefore, while the generic model (8-9) highlights basic principles of the evolution of an excitable nonlinear system, the quasistatic approximation is too limited to be useful for specific systems, and detailed dynamics may be obtained only numerically.

In the following section we turn to the analysis of the full model for mesoscopic collective activity developed by Qin and Sheremet [2022].

## 3. Cross-scale coupling of collective activity

3.1. **The model.** For uni-dimensional excitatory neural fields, the model of collective activity (equations 26 in Qin and Sheremet [2022]) reduces to a system of two coupled evolution equations

$$\frac{\partial}{\partial t} J(x,t) = \Phi H(x,t) - \frac{1}{\tau_j} J(x,t) - N(x,t)\, j_c, \tag{11a}$$

$$\frac{\partial}{\partial t} H(x,t) = \frac{1 - H(x,t)}{\tau_h} - \frac{h_N}{h_0} N(x,t), \tag{11b}$$

where

$$\Phi(x,t) = \epsilon \left( N(x,t) + M_2 \nabla^2 N(x,t) + Q(x,t) \right); \quad N(J) = A \left( \frac{1}{J_c - J} - \frac{1}{J_c} \right). \tag{11c}$$

We refer to Qin and Sheremet [2022] for the details of the derivation of the governing equations. Briefly, the meaning of the notations used here is as follows.

The spatial position vector is denoted by $x = (x_1, x_2, \cdots)$, and $t$ is the time variable.

The state of the neural field is described by two variables: $J(x,t)$, the temperature of the system, and $H(x,t)$, the excitability of the system; The process variables $\Phi(x,t)$, $N(x,t)$, and $Q(x,t)$ represent energy exchanges, internal or with the environment. The variable $\Phi$ denotes internal energy flux, i.e., energy transferred the unit of time by the element of volume of the neural field; $N(x,t)$ represents instantaneous firing rate as the number of spikes in the unit volume of the field and unit of time, normalized by density of cells, and is proportional to the energy released by action potentials; $Q(x,t)$ is the energy flux received from the environment (external forcing) in the unit of time by the unit of volume.

Material parameters describe the geometry and properties of the neural field: $\tau_j$ is the natural decay time constant of membrane kinetic energy; $j_c$ is the sodium channel firing threshold; $\tau_h$ is the natural recovery time constant of excitability (time constant of gating variable $h$ at rest), $\epsilon$ is connectivity strength; and $A$ describes the firing susceptibility of the mean neuron. The connectivity $\epsilon$ is a measure of the efficiency of the transmission of energy fluxes through recurrent connections between neurons. If the normalization of other variables is adequate, the connectivity value should be $O(1)$. The susceptibility $A$ is a measure of the firing likelihood of a cell, given a temperature level $J$. Obviously $A > 0$ is necessary to trigger spikes. Cells do not fire in the limit $A \to 0$, and gradually become more likely to fire when $A$ increases. $J_c$ is the maximum temperature that could be reached by a population, in our context we choose $J_c = j_c$, and for simplicity we normalize the temperature $J$ by $j_c$ such that $j_c = 1$.

3.2. **Quasistatic evolution approximation.** The discussion below is confined to homogeneous forcing and solutions ($\frac{\partial}{\partial x} \equiv 0$, and the dependency on $x$ is dropped). For simplicity, we will keep constant the material parameters $\tau_j$, $\tau_h$ and $j_c$ (time constants and firing threshold) and we assume that the cells fire are mostly the cells in their resting states such that $h_N = h_0$ for analytical solutions. We will account only the effect variations in $\epsilon$ and $A$ (connectivity and susceptibility). If $A = 1$ the firing rate is $N = 1$ if $J = \frac{1}{2}J_c$. This suggests that normal firing activity should be somewhere in the range $A$ between $O(10^{-2})$ to $O(1)$. We will use these ranges in the discussion below.

As before, we introduce the slow time $T = \delta t$, with $\delta \ll 1$ and expand the state variables $p = (J, H)$ as

$$p(t,T) = p_0(T) + \delta p_1(t,T). \tag{12}$$

Assuming again that the external forcing is a large-scale process $Q(T)$, the process variables

$$\Phi(t) = \epsilon\left(Q(T) + N(t,T)\right); \quad N(t) = A\left[\frac{1}{J_c - J(t,T)} - \frac{1}{J_c}\right], \tag{13}$$

may be expanded in a Taylor series around $(J_0, H_0)$,

$$N = N_0(T) + \delta N_1(t,T); \quad N_0(T) = N(J_0(T)); \quad N_1(t,T) = \left.\frac{\partial N}{\partial J}\right|_0 J_1(t,T), \tag{14}$$

$$\Phi = \Phi_0(T) + \delta\Phi_1(t,T); \quad \Phi_0(T) = \epsilon\left(Q(T) + N_0(T)\right); \quad \Phi_1(t,T) = \epsilon N_1(t,T). \tag{15}$$

Substituting into the governing equations (11a-11b) and separating the scales yields, at $O\left(\delta^0\right)$, the algebraic equation for the large-scale state $(J_0, H_0)$

$$\Phi_0 H_0 - N_0 j_c - \frac{1}{\tau_j} J_0 = 0; \quad \frac{1}{\tau_h}\left(1 - H_0\right) - N_0 = 0, \tag{16}$$

and, at order $O\left(\delta^1\right)$, an evolution equation for the mesoscale component $(J_1, H_1)$

$$\left[\frac{d}{dt} + \mathcal{L}\left(J_0, H_0, Q\right)\right] p_1 = -\frac{dp_0}{dt} \tag{17}$$

where

$$\mathcal{L}\left(J_0, H_0, Q\right) = \begin{bmatrix} \epsilon \left.\frac{\partial N}{\partial J}\right|_0 H_0 - \left.\frac{\partial N}{\partial J}\right|_0 j_c - \frac{1}{\tau_j} & \epsilon\left(Q + N_0\right) \\ -\left.\frac{\partial N}{\partial J}\right|_0 & -\frac{1}{\tau_h} \end{bmatrix}. \tag{18}$$

Equations (16), formally identical to the equation for the equilibrium states (Qin and Sheremet, 2022, equation 30) and equivalent to equation (8), describe the slow evolution of large-scale component of collective activity under the slow-varying forcing $Q(T)$. Equations (17- 18) are equivalent to equations (9), and their solution is formally identical to solution (10). The considerations on validity of this approximation (section 2) also apply to equations (16-18).

3.3. **Nonlinear dynamics of collective activity.** The stability analysis of equations (16-18) for constant forcing $Q_0$ is summarized in figure (3). In the space of material parameters $(\epsilon, A)$, a neural field is represented by a fixed point, corresponding to its connectivity and susceptibility (assumed constant). Under constant forcing $Q_0$, one can distinguish two regions: neural fields inside the cusp-like region in figure (3) have triple equilibrium states; fields outside this region have single equilibrium states. In the 1-equilibrium region, the temperature $J$ of the system, the eigen-frequency $\omega$, and the growth rate $\vartheta$ grow counterclockwise around the tip of the cusp (figure 3a-c; see also Qin and Sheremet, 2022). Nearly neutrally-stable equilibrium states are located in a band around the level line $\vartheta = 0$ (blue shaded region in figure 3d).

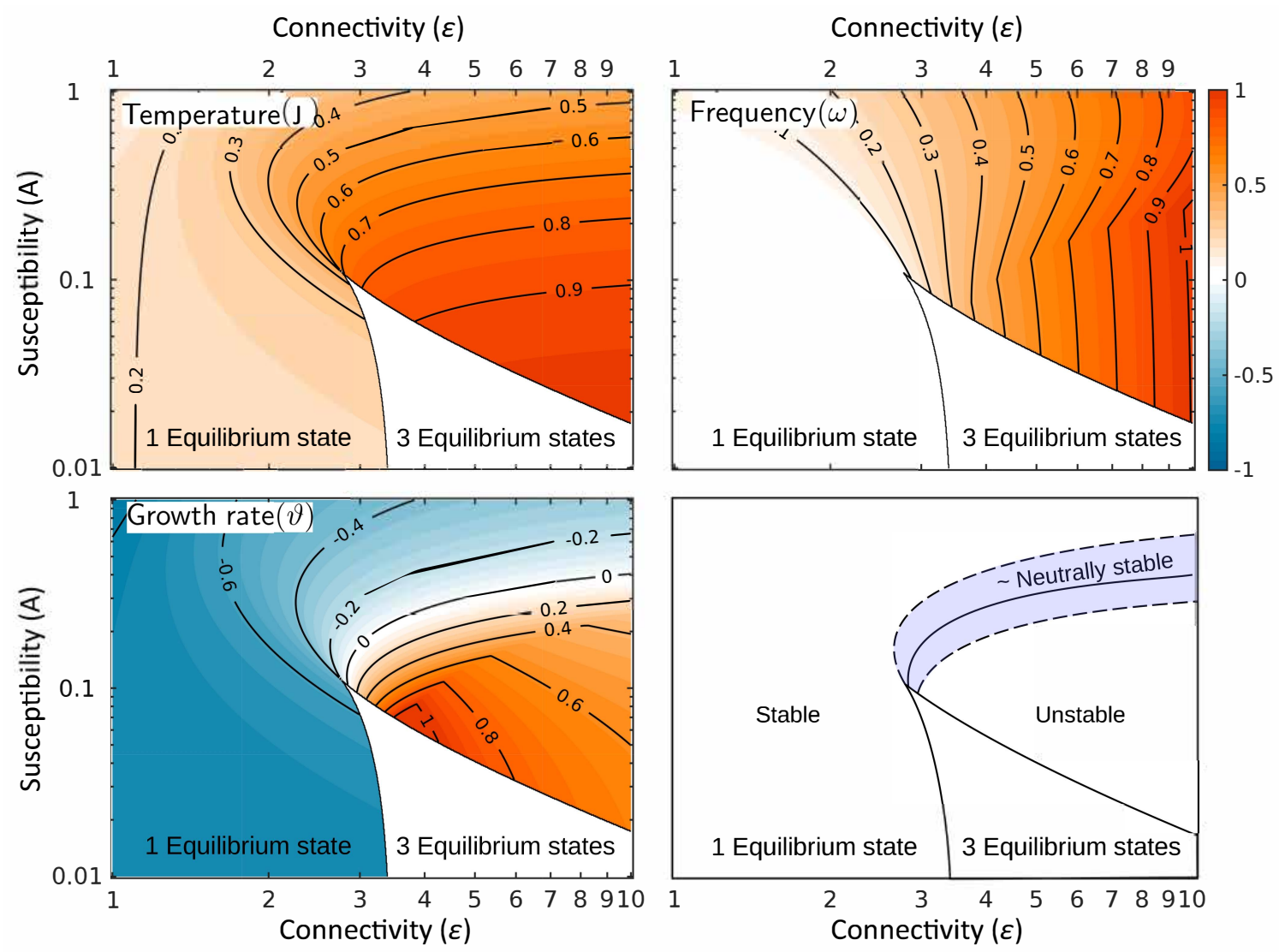

FIGURE 3. Equilibrium states and stability characteristics in the space of material parameters $(\epsilon, A)$. Each point in the $\epsilon$-$A$ plane represents a different "physical" configuration, i.e., a different neural field. a) Field temperature (or kinetic energy density) $J$ at equilibrium states. b) Frequency $\omega$ at equilibrium states. c) Growth rate $\vartheta$ at equilibrium states. d) A schematic map of the stability of 1-equilibrium states. see also Qin and Sheremet, 2022.

In the approximation of quasistatic evolution, the effect of the slowly varying forcing is a change in the position and the shape of the 1- and 3-equilibrium regions (figure 4): as the forcing increases, the 3-equilibrium cusp flattens, and its tip shifts to a lower position in the $(\epsilon, A)$ space. As a consequence, the temperature, eigenfrequency $\omega$, and growth rate $\vartheta$ increase, overall decreasing the stability of any field, regardless of its position in the $(\epsilon, A)$ space (figure 4a). The evolution details of a specific field, however, depend on its position, i.e., its characteristic connectivity and susceptibility values.

A few types of evolution are marked in figure (4) a by the letters A to E. Fields D and E have the simplest evolution: for all $Q(T)$ values shown, D is strongly stable, with mesoscale oscillations suppressed very fast by the strong damping rate; E is strongly unstable, with mesoscale oscillations grow very fast throughout the evolution. Fields A, B, and C are most likely to show persistent mesoscale oscillations modulated by large-scale activity. The path of field A in the state space $(J, H)$ is approximately weakly stable (figure 4a,b,f). Field B transitions between stable and unstable states about halfway through the forcing range shown (figure 4a,c,g). Field C has the most complicated evolution, that includes stable, unstable, and 3-equilibrium states (figure 4a,e,h).

The evolution of fields B and C cannot be described using the linear approximation discussed in section and require an examination of the full nonlinear model 11. However, numerical integration of the nonlinear model (11) shows that linear unstable equilibria are globally stable, surrounded by a stable orbit (figure 4c,g). The transition from stable to unstable equilibria is a therefore bifurcation of the supercritical Hopf type, and the ordering postulated by equation (11) remains valid. The most complicated evolution, field C, exhibits a transition from a single stable equilibrium, to an unstable state surrounded by a stable orbit, but through the 3-equilibrium region (an unstable equilibrium flanked by two stable ones). In fact, the nonlinear model shows a weak supercritical Hopf bifurcation also for path A (figure 4b,f).

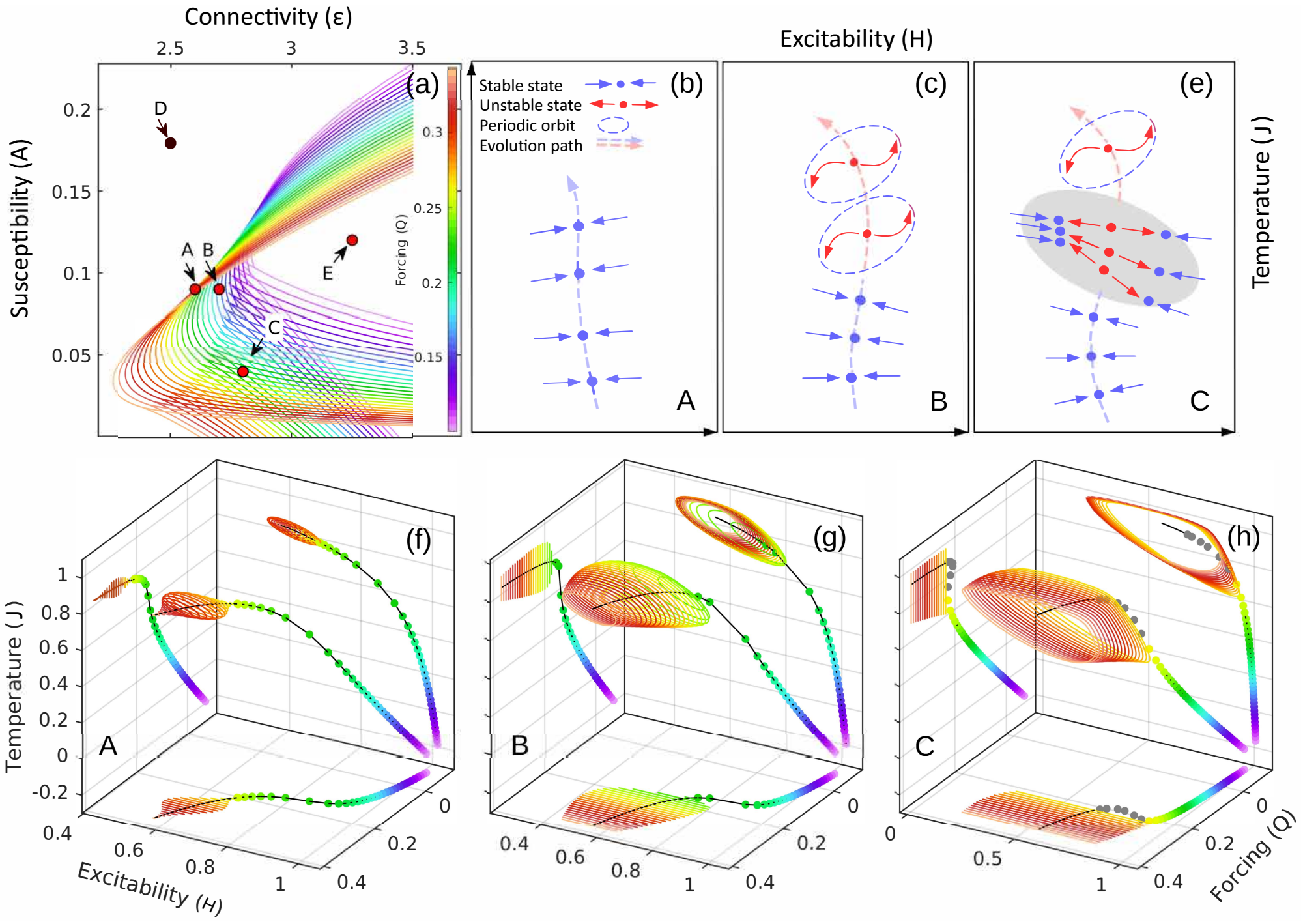

FIGURE 4. Shifting of stability characteristics as a function of forcing $Q$. a) The change of schematic map of the stability, panel d in (3), versus external forcing $Q$. The unstable domain expands toward lower-left domain as $Q$ increases, thus, a "physical" configuration which is initial stable at the lower-left region can be perturbed to generate oscillations by increasing the forcing $Q$. A few types of evolution are marked by the letters A to E. b), c) and e) demonstrate the change of stability for configuration A, B and C respectively by cartoon drawings, while f) h) and h) show the global stable orbits versus forcing $Q$ for configuration A, B and C respectively.

3.4. **Stochastic cross-coupling of collective activity.** The nonlinear evolution (11) enriches the crude linear mechanism of cross-scale coupling of collective activity discussed in section (2). The slow forcing $Q$ changes the stability of the system by modifying the position and shape of critical stability elements in the $(\epsilon, A)$ space. Assume that the forcing cycles periodically between high and low values (e.g., figure 4a), and that the system is in a stable equilibrium region at low forcing values. As the forcing increases, depending on the characteristic values $(\epsilon, A)$ of the field, collective activity may be forced from a stable to an unstable state, where transitional perturbations will excite eigen-oscillations that grow and, given enough time, converge to stable orbits surrounding the unstable state. When the forcing decreases back, the system returns to stable states and eigen-oscillations are suppressed. Thus transient groups of mesoscale collective activity are correlated periodically with high values of forcing.

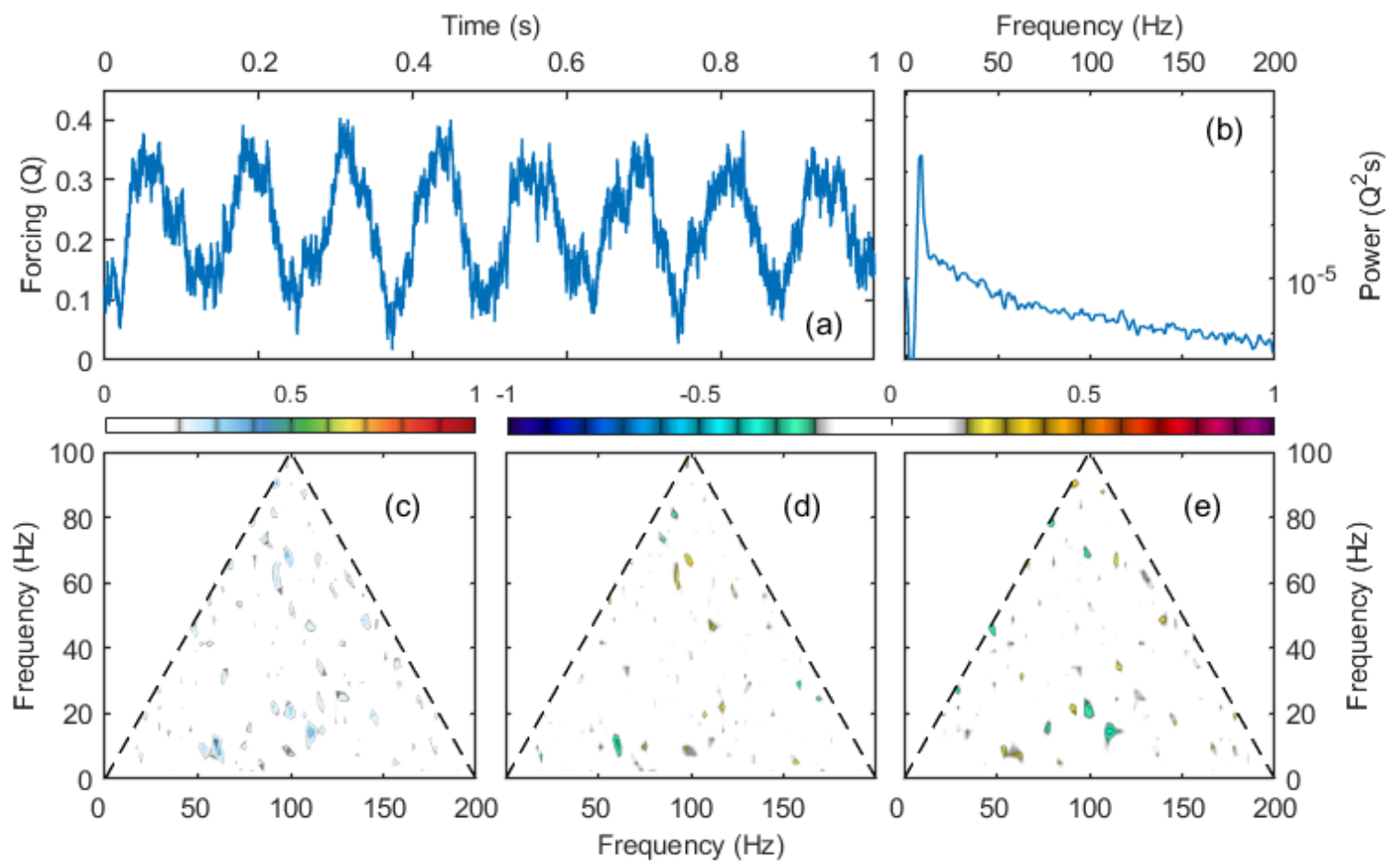

FIGURE 5. A segment of the time series $Q(t)$ used as forcing in both linear and nonlinear numerical simulations. a) A 1 second segment of the forcing time-series; b) Spectral density of $Q(t)$ power; c-e) modulus, real and imaginary part of the $Q$ bispectrum.

Here, we illustrate the similarity between this destabilization process and LFP cross-frequency coupling using the examples of fields A, B, and C identified in figure (4)a. The procedure numerical followed here comprises four steps: 1) represent forcing $Q$ as a stochastic process; 2) integrate the full nonlinear model forced by $Q$; 3) use the time series of $Q$ and collective activity temperature $J$ to construct a "mock" LFP; and 4) apply bispectral analysis to the "mock" LFP for comparison with cross-frequency coupling estimates for rat LFP (figure 2).

The simulated forcing time series $Q$ used in all numerical simulations presented here is shown in figure (5). The time series is a non-skewed, symmetric stochastic process that has a narrow-spectrum Gaussian pink noise $\propto f^{-1.5}$ superposed on a monochromatic oscillation of 8 Hz (arbitrary value, chosen for demonstration purpose). Through construction, the forcing time series does not contain any cross-frequency coupling, as illustrated by the bispectral analysis: the modulus (the real and the imaginary part) of the bispectrum is statistically zero (figure 5c-e).

Together with the forcing time series $Q$, the state variables $(J, H)(t)$ fully describe the evolution of the system. Although, as stressed above, equations (11) provide a crude representation of brain activity, it is of interest to seek some comparison with LFP recordings from rat hippocampus. The time series $Q$ is a representation of externally forced synaptic excitation of the local network and the temperature $J$ represents somatic collective activity in response of firing events. For example, if local network were CA1 region in the hippocampus, $Q$ might represent excitation incoming though Schaffer collateral path, and $J$ might represents current source in the vicinity of stratum pyramidale. Both these signals contribute to, but are not the only components of an LFP recording in the rat hippocampus, which combines synaptic and somatic activities, spiking and hyperpolarization activities, and a number of other ionic processes [Buzsáki et al., 2012]. Because the model discussed here does not account for many of these processes, we construct here a "mock" LFP by the simple superposition

$$\text{LFP} \sim -(Q + J)\,. \tag{19}$$

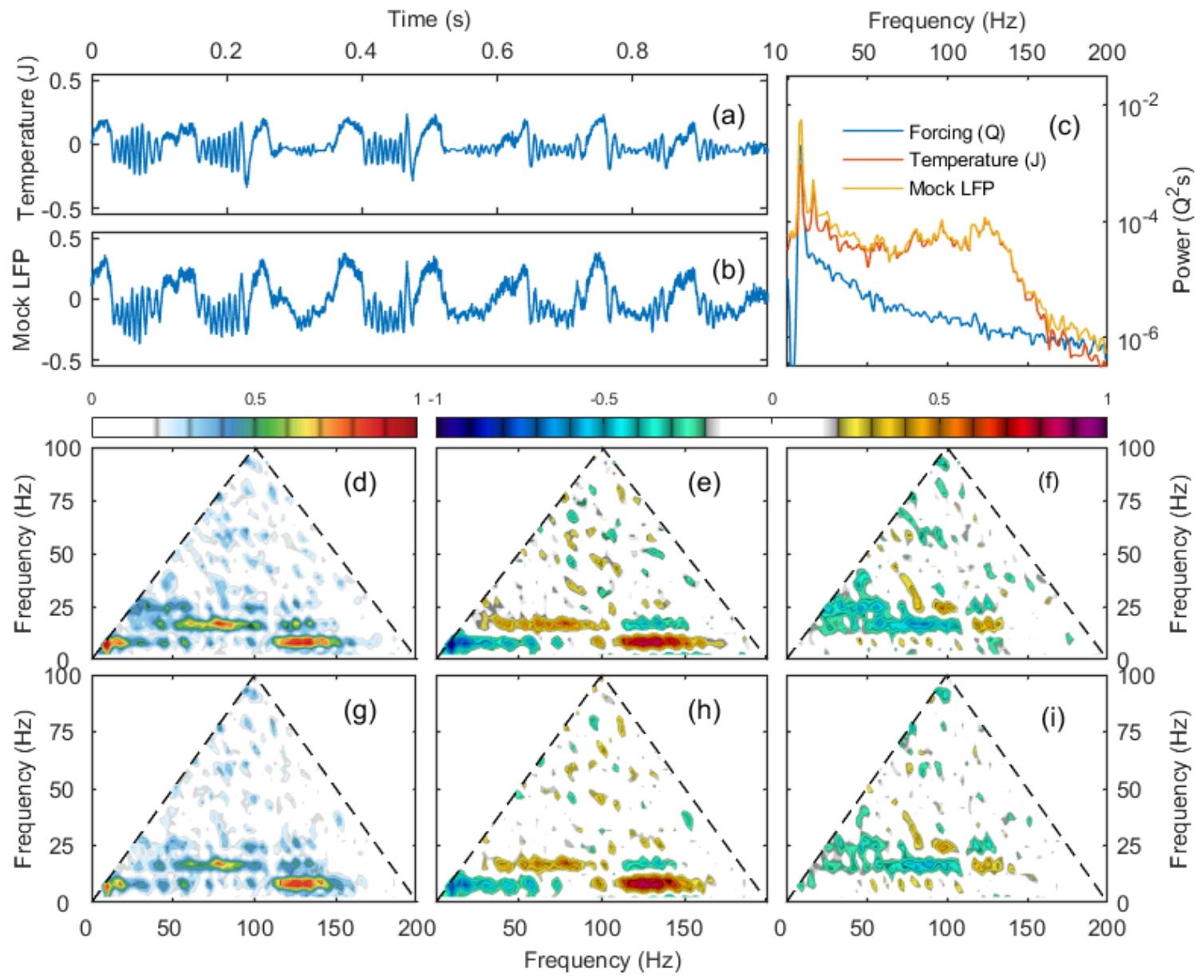

FIGURE 6. A numerical simulation of the linearized system 16-17 for neural field $A$ (figure 4), nearly stable under the range of forcing $Q$ defined in figure 5. Segments of time series of a) temperature response of the field, and b) mock LFP – a model of what an LFP trace might look like, see text for explanations; c) spectrum density of power for the forcing $Q(t)$, field temperature $J(t)$, and mock LFP; d-f) temperature bispectrum: modulus, real, and imaginary part; g-i) fake LFP bispectrum: modulus, real, and imaginary part. Note that the bispectra demonstrate two types of cross-frequency patterns that may correspond to the theta harmonics (slow gamma) and theta-gamma coupling.

3.5. **Numerical results.** Examples of numerical simulations are shown in figures (7-9). All simulations exhibit qualitatively similar behavior. We use the system temperature $J$ illustrates the raw response of the system; the "mock" LFP suggests an LFP trace might look like. The response of the system ($J$) shows two scales of activity (spectra in panels c in the figures): a large scale component at 8 Hz, matching the forcing, and groups of mesoscale activity (eigen-oscillations or limit orbits) between 75 and 150 Hz, phase-correlated to the peaks of the forcing and troughs of the mock LFP.

Bispectral analysis shows two types of cross-frequency patterns, one related to deviations of the large-scale response of the system from the sinusoidal shape of the forcing $Q$; the other is related to the cross-scale coupling between large scale and mesoscale activity.

The deformation of the large-scale time series is reflected in the appearance of phase-coupled harmonics of the 8 Hz forcing oscillation. Although the forcing is free of any cross-frequency coupling (figure 5), the large-scale collective activity response shows significant skewness and asymmetry, obvious in the time series (panels a-b), and also detectable as harmonics of the 8-Hz peak in the spectra and bispectra of temperature and "mock" LFP (panels c, and d-i) and (forcing spectrum contains no harmonics; panels c). The source of skewness seems to be the "trapping" of the system by groups of mesoscale activity – either stable orbits in the nonlinear cases (figures 7- 9), or eigen-oscillations in the linear case (figure 6).

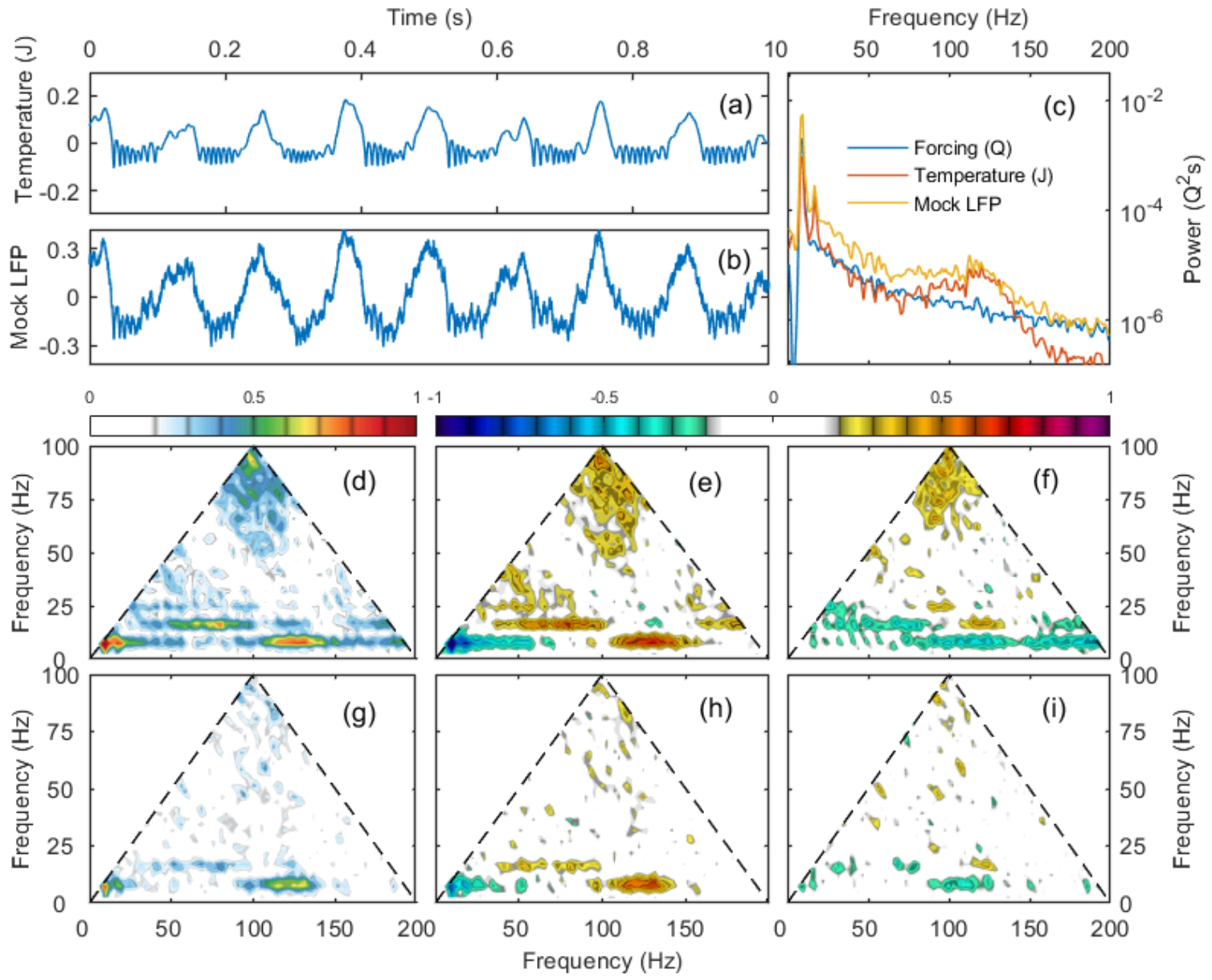

FIGURE 7. A numerical simulation of the nonlinear equations 11a-11b for neural field $A$ (figure 4), nearly stable under the range of forcing $Q$ defined in figure 5. Time series of (a) temperature and (b) mock LFP. c) Spectral densities of temperature, mock LFP and forcing. Modulus, real, and imaginary part of the bispectrum of temperature (d-f) and mock LFP (g-i). Compare with figure 6 (same organization as figure 6).

The cross-scale coupling between large-scale and mesoscale components reflects the correlation between mesoscale groups and peaks of forcing (e.g., time series, panels a-b). In the bispectrum, this is expresses as non-zero bands spanning the mesoscale frequencies (approx 100 to 150 Hz) on the horizontal axis, panels (d-i).

While in the case of field A mesoscale oscillations are mainly driven by the noise in the forcing time series, and amplified in the weakly unstable region, in the evolution of fields B and C the system instability plays a crucial role (the linear approximation is invalid), and mesoscale activity is defined largely by the characteristics of the limit orbits. Comparing the evolution of fields B and C (figures 4g-h, 8, and 9), the slower oscillations associated with larger limit orbits of C lead slower mesoscale oscillations, approximately 75 to 125 Hz.

As an example of evolution that stays close to stable equilibria (figure 4, panels b and f), field A may be used to illustrate the limitations the linear quasistatic approximation. Although the linear and nonlinear solutions are similar (compare figures 6 and 7), in the small time span in which the system is near weakly unstable states (see limit orbits in figure 4f) the behavior is quite different. In the linear system the eigen-oscillations grow unchecked, resulting in large mesoscale amplitudes and power (figure 6a-c) likely breaking the ordering in equation 6). In the nonlinear simulations, mesoscale amplitudes (power) are is bounded by the convergence to the limit orbits.

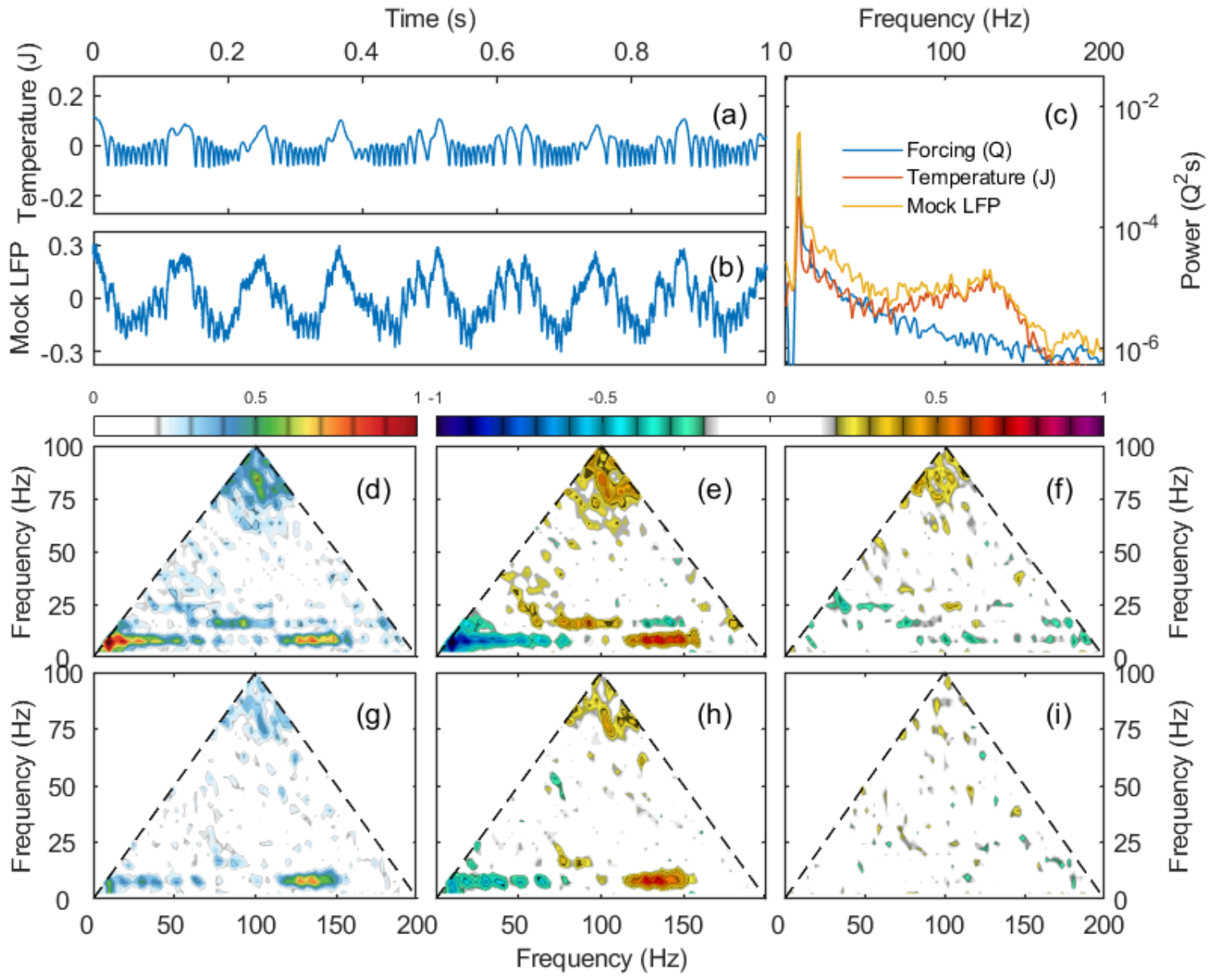

FIGURE 8. A numerical simulation of the nonlinear equations 11a-11b for neural field $B$ (figure 4), which becomes unstable at high values of the forcing $Q$ defined in figure 5 (same organization as figure 6).

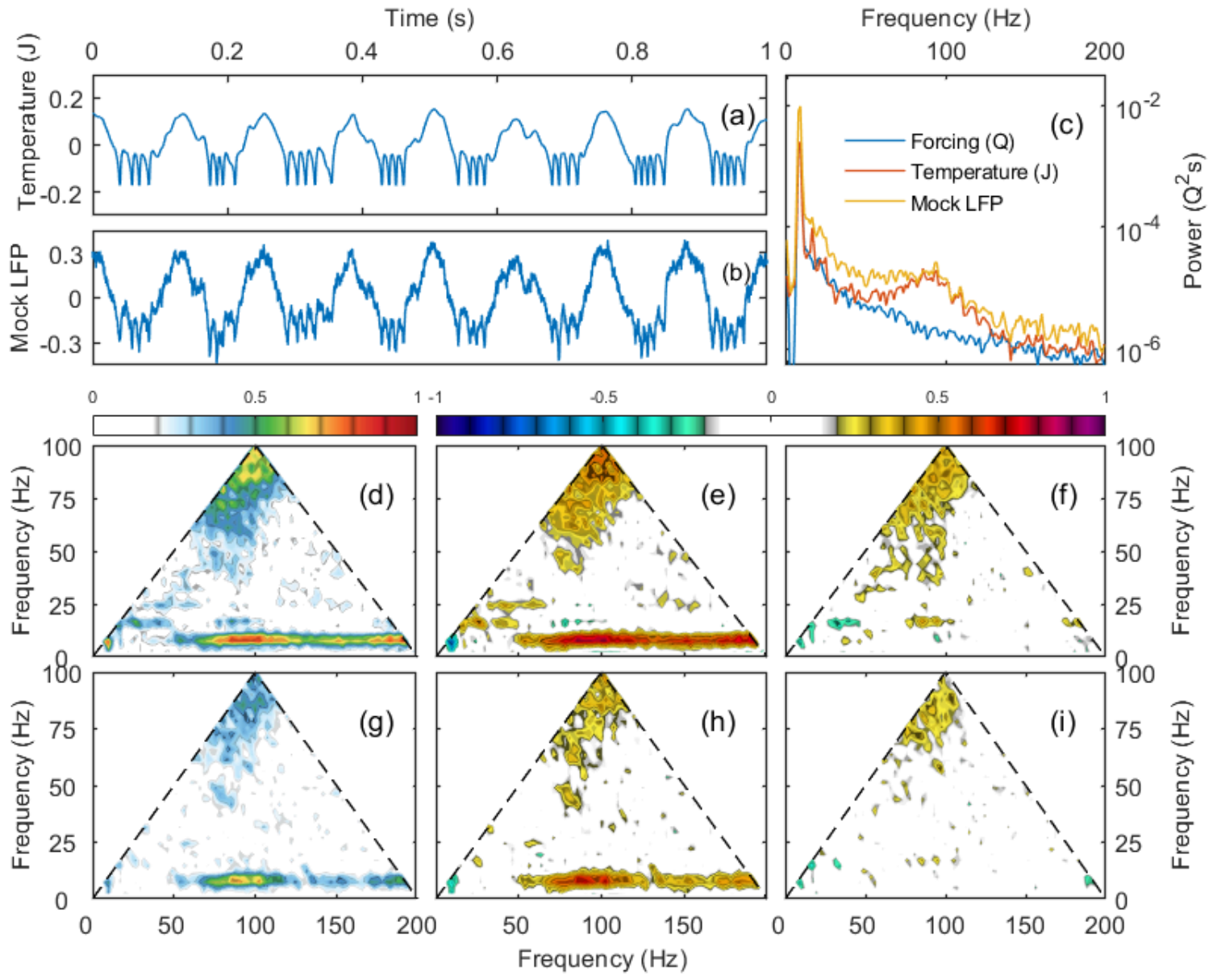

FIGURE 9. A numerical simulation of the nonlinear equations 11a-11b for neural field $C$ (figure 4), which cycles through all stability regions identified by the linear analysis (see figures 3 and4) under the range of forcing $Q$ defined in figure 5 (same organization as figure 6).

## 4. SUMMARY AND DISCUSSION

Theta-gamma coupling has been explained as coupled activity of specialized cell pairs (excitatory-inhibitory neurons), or of specialized cells (fast bursting neurons). This explanation, however, does not address observations of theta-gamma coupling in non-specialized excitatory networks whose individual cells connect and behave randomly. Its most intriguing feature, however, is that the underlying mechanisms are uniquely specific to the brain. In fact, frequency coupling (in general, cross-scale coupling) is a phenomenon encountered in all physics of large, multi-scale systems. In these systems, cross-scale coupling is generated by nonlinear interaction between the system components. Because the brain is also a nonlinear, multi-scale physical system, a similar mechanism must be active in the brain, possibly manifesting itself in LFP recordings as theta-gamma coupling.

In this study, we hypothesize that cross-frequency coupling is a consequence of nonlinear dynamics of collective activity, understood as spatio-temporal activity patterns in the cortex (e.g., waves). *Remarkably, this mechanism does not require specialized cells or cell pairs; its sole ingredient is nonlinearity. It is therefore universal, and must be active also in neural fields of any composition.*

The universality of the nonlinear coupling principle is demonstrated first on a generic nonlinear equation, and confirmed using the linearized governing equations of excitatory collective activity [Qin and Sheremet, 2022], which exhibit identical properties.

A qualitative insight into the dynamics of theta-gamma coupling may be obtained by taking advantage of the power mismatch between the two oscillations to linearize the system of equations. The linearized equations represent theta-gamma coupling as a forcing "cascade": theta arises as a direct response to large-scale external forcing, while gamma oscillations are driven jointly by the external forcing and theta, the latter through the nonlinear term of the equations.

Although the quasistatic approximation has limited applicability, it does provide insights into the role played by excitability, forcing, and nonlinearity in the evolution of the neural field.

The excitable nature of the system is expressed in the balance of the energy fluxes $\Phi$ recaptured from action potentials and the natural temperature decay (the $-\frac{1}{\tau_j}J$ term in equation (11a)). The external forcing plays the basic role of maintaining activity in the system. The Qin and Sheremet [2022] model is fundamentally a dissipative system. Without an external supply of energy, the field temperature (density of kinetic energy carried by sodium ion currents) decays naturally, reducing firing rates and slowly cooling off the system. Self-sustained activity without external input is theoretically possible at high temperatures, if the energy recaptured from action potential approximately balances the decay rate, but the system is likely unstable.

Excitability and external forcing could be described as combining to determine the nature and behavior of mesoscale perturbations (oscillatory, decaying or growing), by creating a rich pattern of equilibria with various stability characteristics. Nonlinearity, here represented as forcing by large scale activity, governs the generation of mesoscale activity.

Because quasistatic solutions are not valid for unstable states, the properties of the solutions are further examined by numerically integrating the full Qin and Sheremet [2022] equations. Numerical simulation scenarios illustrate the evolution of the system for different stability cases. Nearly-neutrally stable evolution (case A) is largely linear, with nonlinearity making a brief appearance at unstable states to control the amplitude of the mesoscale process through the presence of limit orbits. If the evolution path goes through unstable states (B and C), nonlinear evolution is stronger, and dominates at states with higher energy.

Bispectral analysis of the numerical simulations show cross-frequency patterns similar to observations. As in LFP recordings, high-power large scale oscillations exhibit strong statistical deviations from sinusoidal shapes (skewness, non-zero real part of bispectrum), and are accompanied by phase-coupled mesoscale activity, the amplitude modulation. Interestingly, the skewness of the large-scale process is not introduced by the external forcing, rather, it may be interpreted as a small feed-back modulation of the large-scale process by the mesoscopic process – an effect that was neglected in in the "forcing cascade" representation.

Our nonlinear numerical simulations do indeed show that nonlinearity can generate theta-gamma coupling in excitatory-only fields. However, it is important to note that the simulations do not contain any a priori information of what the frequency band of the gamma oscillation should be. Nonlinearity transfer the power from theta to some higher frequency, and generates coupling. However, the location and nature (oscillatory, decaying or growing) of receiving high-frequency band is determined by combined effect of excitability and external forcing. Perhaps unsurprisingly, inhibition is an essential factor in the determination of these properties. Indeed, in equations (11), letting the refractoriness (an inhibition mechanism) time constant $\tau_h \to 0$ implies $H \to 1$, and equation (11b) is disabled. We conclude that, the nature gamma oscillations is controlled by inhibition. Therefore, the presence of inhibitory cells is likely to produce a richer nonlinear behavior.

We conclude this discussion with a look past the simple scenario of two-frequency interaction. After all, description of gamma as a single frequency is rather crude, given the fairly wide band associated with it (e.g., 65–130 Hz Sheremet et al., 2020); and our simulations also distribute the forcing energy coming from the theta band over a rather wide band (100-150 Hz, e.g., figures 7-9). In general, nonlinear redistribution of energy across scales may be regarded as a cross-scale spectral energy flux, which typically flows away from the input scale and toward the dissipation scale. If input and dissipation occupy scales that are far enough apart (e.g., input at large scales, and dissipation at small scales), and if nonlinear interaction is local in scale (efficient only between neighboring scales) there exists a spectral window where only nonlinearity is active. This window is called inertial spectral window. At equilibrium, the dissipation rate must match the input rate, which implied that the cross-scale flux in the inertial window is constant. Because all scales in the inertial window share the same physics (nonlinearity), the spectral distribution in the inertial window must be scale-invariant (power law). These spectra are called Kolmogorov-Zakharov spectra [Kolmogorov, 1941, Zakharov et al., 1992a]. The energy flux in the inertial window is called the energy cascade.

The concept of the cross-scale energy flux may be a useful tool for understanding neural collective activity. In the cortex, the input scale for collective activity is likely in the theta scale. Judging by figure 7b in Qin and Sheremet [2022], dissipation rates increase sharply at high-frequencies (wave numbers). This suggests that an inertial range might exist in the gamma band. Scale invariant spectra are typically observed in the gamma band at low power (low field temperature; e.g., figure 1), while high power spectra (high input rate) are no longer self similar, and exhibit a broad gamma peak (figure 2). Self-similar, low power spectra might indicate weak nonlinearity and energy transfers (cross-scale fluxes) in the gamma band (inertial window). A possible explanation for the departure of high power spectra from scale invariance is that the spectral flux is no longer constant, decreasing at higher frequencies. This would cause a "bottleneck" at some high frequency, and an accumulation of power at frequencies just below it.

Finally we stress that, the goal of this study was to demonstrate that nonlinearity is a fundamental mechanism for cross-scale energy transfers and cross-frequency coupling, and that it likely plays an important role in the generation of gamma oscillation. While the Qin and Sheremet [2022] model is quite general and will hopefully prove useful for applications to brain activity, *it is not yet a model of the brain*, because it lacks quite a few capabilities (e.g., calcium channels).

**Compliance with Ethical Standards.** Disclosure of potential conflicts of interest: The authors declare no conflict of interest in the research described in this paper.

Research involving Human Participants and/or Animals: No human or animals were involved in the research described in this paper.

Informed consent: Not applicable.

## Appendix A. The bispectrum

**Stochastic Fourier analysis.** If $p(t)$ is a stochastic process stationary in the relevant statistics sampled at $t_j =$, the discrete Fourier direct and inverse transform is defined as [e.g., Weaver, 1989]

$$P_n = \sum_{j=0}^{N-1} p_j \exp\left(-2\pi f_n t_j\right); \quad p_j = \frac{1}{N}\sum_{n=0}^{N-1} P_n \exp\left(2\pi f_n t_j\right); \tag{20}$$

where the time and frequency grids are $t_j = j\Delta t$ and $f_n = n\Delta f$, with $\Delta f = \frac{1}{N\Delta t}$, $j, n = 0, 2, \ldots, N-1$, and $N$ is the number of points in the time domain.

The discrete cross-spectrum and cross-bispectrum estimators for three stochastic processes $p$, $q$, and $r$, are

$$S_n^{pq} = S^{pq}\left(f_n\right) = \langle P_n Q_n^* \rangle, \tag{21a}$$

$$B_{mn}^{pqr} = B^{pqr}\left(f_m, f_n\right) = \langle P_m Q_n R_{m+n}^* \rangle. \tag{21b}$$

where $P$, $Q$, and $R$ denote the DFTs, angular brackets represent the ensemble average, and the asterisk denotes complex conjugation. The cross-spectra describe the frequency distribution of the variance of processes $p$ and $q$, i.e., a complete characterization of the average linear structure of the Fourier representation. For two distinct processes $p$ and $q$, the cross-spectrum provides a measure of their correlation, typically represented as the coherence and phase lag

$$C_n^{pq} = \frac{S_n^{pq}}{\sqrt{S_n^{pp} S_n^{qq}}}, \text{ and } \Theta_n^{pq} = \arg S_n^{pq}. \tag{22}$$

For a single stochastic process the cross-bispectrum is generally referred to as bispectrum, $B_{mn}^{ppp} \equiv B_{mn}^{p}$ (the superscript may be dropped if there is no ambiguity). A variance factor is typically used to eliminate the variance bias in the magnitude of the bispectrum: e.g., the normalization

$$b_{mn}^{p} = \frac{B_{n,m}^{p}}{\left(\left\langle |P_n P_m|^2 \right\rangle \left\langle |P_{n+m}|^2 \right\rangle\right)^{1/2}}. \tag{23}$$

insures that $\left|b_{mn}^{p}\right| \leq 1$ [Haubrich and MacKenzie, 1965, Elgar and Guza, 1985]. The bispectrum arises naturally as a measure phase coupling between Fourier components of time series $p$: for a Gaussian time series, $\left|b_{mn}^{p}\right|$is statistically zero, while $\Re\left\{b_{mn}^{p}\right\} \Im\left\{b_{mn}^{p}\right\}$ are related to the skewness and asymmetry of $p_j$ [Haubrich and MacKenzie, 1965, Masuda and Kuo, 1981].

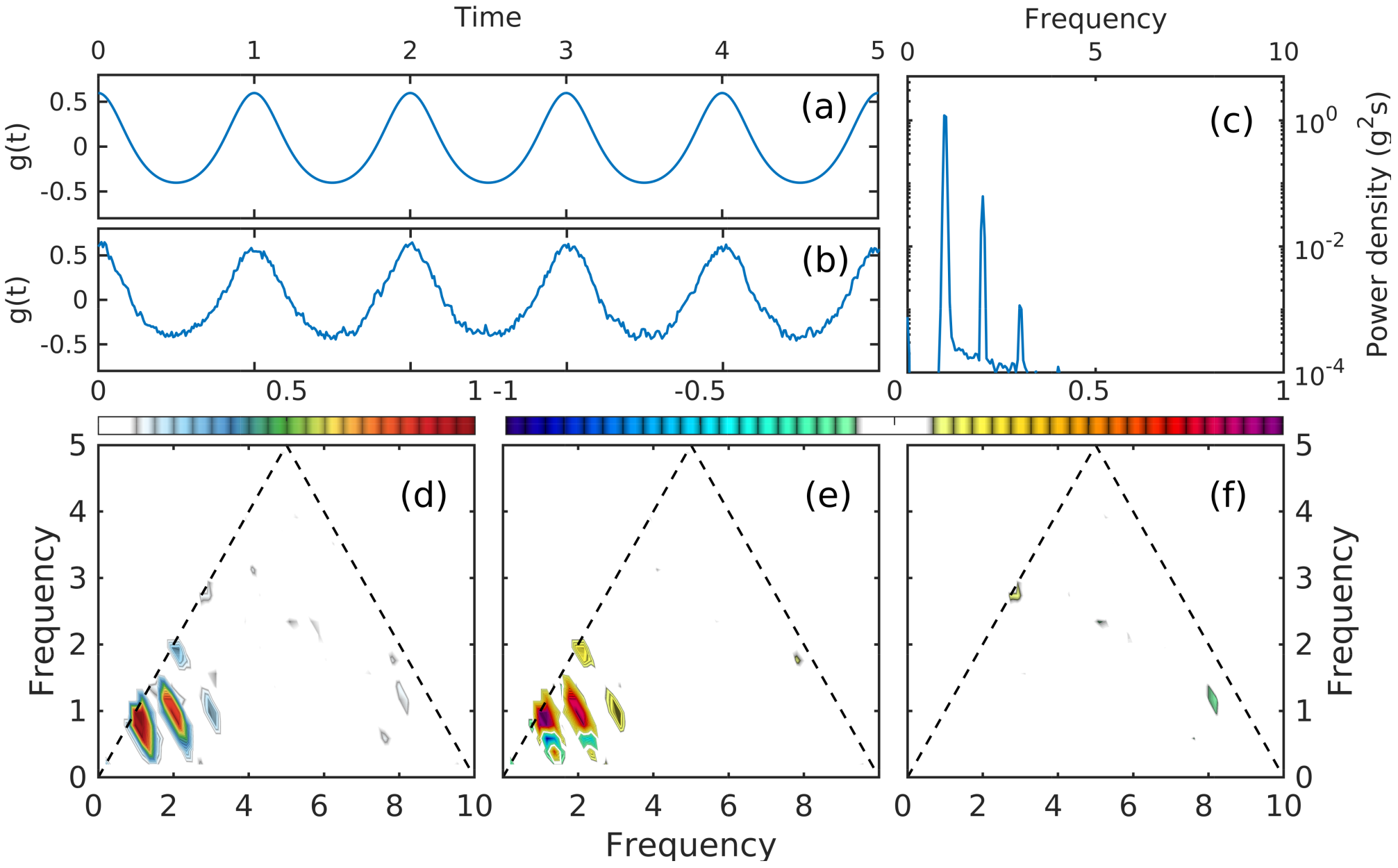

FIGURE 10. Schematic of the interpretation of the bispectrum. a) Due to the symmetries of the bispectrum [e.g., Sheremet et al., 2020] the principal (non-redundant) domain of $b^p_{mn}$ is octant 1 in the $(f, f)$ plane (shown here bounded by $f_N$ (for plotting convenience, the frequency axes are not shown equal). A region in the plane $(f, f)$ such that $b^p_{mn} > 0$ (circle) indicates cross-frequency coupling between the frequencies corresponding to the orthogonal projections of the region on the two axes (blue, centered at $f_m$ and $f_n$), and the projection of the region on the horizontal axis along the second diagonal (red, centered at $f_m + f_n$). b) Smooth time series, positively skewed ajnd symmetric. c) "Noisified" time series: weak Gaussian pink noise was added to the time series in (b), to enhance the bispectral peaks. d) Spectrum of time series (c). e-f) Components of the bispectrum of time series (c): modulus $\left|b^p_{mn}\right|$, real part $\Re\left\{b^p_{mn}\right\}$ (measure of skewness) and imaginary part $\Im\left\{b^p_{mn}\right\}$ (measure of asymmetry) bispectrum . The peaks in $\left|b^p_{mn}\right|$ indicate cross-frequency coupling between the fundamental (unit frequency) and its harmonics; they are prominent and positive in $\Re\left\{b^p_{mn}\right\}$, indicating that the coupling is associated with positive skewness; $\Im\left\{b^p_{mn}\right\}$ is statistically zero, indicating that the time series is symmetric.

The interpretation of the bispectral components is illustrated in figure 10. he use of cross-bispectra was first proposed to describe the third order statistics of ocean waves ([Hasselmann et al., 1963]. Bispectral analysis has a wide application in nonlinear systems, ranging from water waves, large scale interplanetary scintillation, plasma turbulence, to small scale pitch detection, image reconstruction, and machine fault diagnosis (see, e.g., Elgar and Guza, 1985, Spicher et al., 2015, Itoh et al., 2017, Chiang and Nikias, 1990. Applied to EEG recordings [e.g., Kleiner et al., 1969, Isler et al., 2008], bispectral analysis has been used to connect cross-frequency coupling to physiological states [Bullock et al., 1997, Andrzejak et al., 2001]. and to investigate the evolution of the nonlinear character of the hippocampal LFP as a function of rat running speed. Sheremet et al. [2016]. An in-depth discussion of the bispectral estimator in relation with the nonlinear coupling estimators popular in neuroscience can be found in Kovach et al. [2018].